\documentstyle[12pt]{article}
\begin{document}
\renewcommand{\theequation}{\thesection.\arabic{equation}}
\newcommand{\beq}{\begin{equation}}
\newcommand{\beqa}{\begin{eqnarray}}
\newcommand{\eeq}{\end{equation}}
\newcommand{\eeqa}{\end{eqnarray}}
\newcommand{\p}{\partial}
\newcommand{\bp}{\bar{\partial}}
\newcommand{\tsm}{ topological $\sigma$-model }
\newcommand{\tsms}{ topological $\sigma$-models }
\newcommand{\sm}{ $\sigma$-model }
\newcommand{\sms}{ $\sigma$-models }
\newcommand{\sigtag}{semi-infinite cigar-like target space }
\newcommand{\bhm}{black hole metric }
\newcommand{\mv}{\mid v \mid}
\newcommand{\mw}{\mid w \mid}
\newcommand{\mr}{\mid \rho \mid}
\newcommand{\dm}{\partial_{\mu}}
\newcommand{\dn}{\partial_{\nu}}
\newcommand{\s}{\sigma}
\newcommand{\la}{\lambda}
\newcommand{\e}{\varepsilon}
\newcommand{\al}{\alpha}
\newcommand{\bt}{\beta}
\newcommand{\r}{\rho}
\newcommand{\wi}{w^{i}}
\newcommand{\wj}{w^{j}}
\newcommand{\lmi}{\lambda^{\mu i}}
\newcommand{\lap}{\lambda^{++}}
\newcommand{\lam}{\lambda^{--}}
\newcommand{\lnj}{\lambda^{\nu j}}
\newcommand{\hi}{\chi^{i}}
\newcommand{\hj}{\chi^{j}}
\newcommand{\hk}{\chi^{k}}
\newcommand{\br}{\bar{\rho}}
\newcommand{\bw}{\bar{w}}
\newcommand{\emn}{\varepsilon_{\mu\nu}}
\newcommand{\g}{g_{ij}}
\newcommand{\J}{J_{ij}}
\newcommand{\intx}{\int d^{2}x}
\newcommand{\II}{$I\bar{I}\;$}
\newcommand{\bi}{$\bar{I}\;$}
\title{The Broken Phase of the  Topological \\ Sigma model}
\author{A.V.Yung \thanks{Permanent address: St. Petersburg Nuclear
Physics Institute, Gatchina, St. Petersburg 188350, Russia.}
\thanks{Research supported in part by the Russian Foundation
for Fundamental Studies under grant no. 93-02-3148}\\
University of Wales, Swansea SA2 8PP, UK}
\date{January 1994}
\maketitle
\begin{abstract}
The \tsm with the \sigtag (black hole geometry) is considered.
The model is shown to possess unsuppressed instantons. The
noncompactness of the moduli space of these instantons is responsible
for an unusual physics. There is a stable vacuum state in which the vacuum
energy is zero, correlation functions are  numbers thus the
   model is in the topological phase. However,
there are other vacuum states in which  correlation functions show
 the coordinate dependence. The estimation of the  vacuum energy
indicates that it is  nonzero.
 These states
are interpreted as the ones with broken BRST-symmetry.
\end{abstract}
\vspace{1cm}
SWAT 94/22 \\
hep-th 9401???
\newpage

\section{Introduction}

Topological Field Theories (TFT) received  considerable attention during
recent years. After the discovery of the topological Yang-Mills theory
in four dimensions \cite{WYM} and topological $\s$ models in two
dimensions \cite{Wsm} a large class of 2D TFT have been studied, including
conformal TFT and their deformations. Particularly interesting
results have been obtained for \tsms formulated on Calabi-Yau spaces
and for their relationship to N=2 superconformal theories in the
 Landau-Ginzburg
formulation (see  \cite{CY/LG} and references therein).

The problem we are attempting to address in this paper is how TFT could
serve  physics. The puzzle stems from the very notion of TFT:
they have no physical degrees of freedom. That means that they have no
propagating particles; the configuration space in these theories is a
finite dimensional one.

The most popular approach to TFT nowadays is the topological string theory.
It was proven \cite{Wgr,Li} that the non-critical
 string theory at low dimensions
( with central charge of matter $c<1$ ) is equivalent to topological minimal
matter \cite{Eguchi} coupled to the topological 2D gravity \cite{Wgr}.
However, it is not a surprising that low-dimensional string turns
out to be a topological theory, because at $c<1$ string has no physical
degrees of freedom. The remaining open problem is whether TFT could be
relevant to the  string theory at $c\geq 1$. In particular, the case
$c=1$ is under intensive study \cite{Bersh,MV}.

Another motivation to study TFT has been put forward by E. Witten in his
original papers \cite{WYM,Wsm}. In the framework of the field theory
one can try to construct the theory of a physical gravity as a broken
phase of the  topological gravity. In this setup one could hope to
arrive at the theory of gravity with good ultraviolet properties, at
least renormalizability. The idea is similar to that in the electroweak
theory: the spontaneous breakdown of the gauge symmetry does not affect
ultraviolet properties of the model.

However, little is known about the possibility of the spontaneous breakdown
of the BRST-symmetry in TFT. One known example is the topological quantum
mechanics with double-well potential studied in ref.\cite{brQ} (see \cite{rev}
for a review). The mechanism of the BRST-symmetry breaking in that case is
essentially the same as in the N=1 SUSY quantum mechanics considered
 by E. Witten
long ago \cite{WSUSY}. Namely, the supersymmetry is dynamically broken
by instantons. The main reason for this phenomenon is the noncompactness
of the moduli space of instantons, which is a real line (rather than circle)
in the case at hand.

In TFT this breakdown manifest itself as a nonzero expectation value of
BRST-exact operators. This is the case for the quantum mechanical example
mentioned above, nevertheless such fundamental properties of TFT as
coordinate and coupling constant independence of correlation functions
of observables still remains untouched \cite{brQ}. Therefore, one cannot
speak about ``liberating''  any physical degree of freedom in that case.

In this paper we consider the 2D \tsm \cite{Wsm} with the \sigtag. This
geometry of the target space arises naturally \cite{Wbh} as a d=2
 black hole solution
in the string theory associated with $SL(2,R)/U(1)$ gauged WZW model
\cite{GWZW} ($d$ is the dimension of the target space) \footnote{Note, however,
that we do not consider the topological version of the $SL(2,R)/U(1)$
WZW theory here. The latter model would be a conformal TFT. The $\s$
model we study here is not a conformal one due to the absence of the
dilaton term in the action.}

In  one of our previous papers \cite{Ybh} it is shown that the black
hole model possesses world-sheet instantons, although the target space of the
model is not compact. Here we discuss this point in some detail in the
context of the  \tsm with \sigtag . It follows from the standard topological
reasoning that $d=2$ \sms with compact target spaces have instantons, while
\sms with noncompact ones do not. We show that the case of the semi-infinite
cigar-like \sm is on the ``borderline'' between compact and noncompact cases
and needs more careful regularization. Our result (which is one of the
main results of this paper) is that the model at hand
does have nonsuppressed instantons.

The moduli space of these instantons turns out to be noncompact. This
ensures the divergence of the integrals over the moduli space for the
correlation functions of observables. These divergences provide the
general reason for the BRST-symmetry breakdown in the model at hand.

The physics in the model turns out to depend on the vacuum expectation
value (VEV) of the \sm boson field $\mv$. We show that  the true topological
phase with zero vacuum energy and coordinate independent correlation
functions is realized only for
 $\mv \rightarrow \infty$. Instead, for finite $\mv$
 correlation functions of
observables show the  coordinate dependence. Moreover, we study
anti-instantons  in the model at hand and show that they produce
 nonvanishing effects. Actually they generate a true interaction
in the model which involves a coupling constant dependence. A preliminary
study of the vacuum energy indicates that it is nonzero.
Our interpretation of these results is that there are degenerative
vacuum states at finite $\mv$ for which
  the BRST-symmetry is broken and some physical
degrees of freedom are ``liberated''. Although these states are unstable
we present qualitative arguments that their lifetime could be infinitely large.

The plan of the paper is as follows. In Sec. 2 we review briefly general
properties of $d=2$ \tsms . In Sec. 3 we consider the \sm with \bhm  and
show that it has world-sheet instantons. In Sec. 4 we study correlation
functions of observables and show their coordinate dependence for finite
$\mv$. In Sec. 5
 we develop  the effective Lagrangian technique
for anti-instanton effects and in Sec. 6 we estimate the vacuum energy.
 Sec. 7 contains our final discussion.

\section{$d=2$ \tsms .}

Here we review some basic properties of $d=2$ \tsms \cite{Wsm,Wgr}. The action
of the model on the $d=2$ Kahler target manifold reads \cite{Wsm}
\beqa
S & = & \frac{r^{2}}{\pi}\intx
\left[
\frac{1}{2}\g (w)\dm w^{i}\dm w^{j}- \frac{1}{2}\J (w)\emn\dm w^{i}\dn w^{j}
\right. \\ \nonumber
 & - & \g \lmi D_{\mu}\hi - \frac{1}{8}R_{ij,kl}\hi\hj\lambda^{\mu i}
\left.
\lambda_{\mu}^{j}
\right].
\eeqa
Here $\mu ,\nu =1,2$ are world-sheet indices, while $i,j=1,2$ are target
space ones. For simplicity the world-sheet is considered to be a complex plane
with Euclidean flat metric $h_{\mu,\nu}=\delta_{\mu,\nu}$ and complex
structure $\emn$, $\varepsilon_{1,2}=1$, while $\g$ and $\J$ are the metric
and the complex structure of the target space. $D_{\mu}$ is the covariant
derivative with respect to the connection on the target manifold
$\Gamma^{i}_{kl}$
\beq
D_{\mu}A^{i}=\dm A^{i}+ \dm w^{k}\Gamma^{i}_{kl}A^{l},
\eeq
while $R_{ij,kl}$ is the target space curvature tensor
\beq
R^{i}_{j,kl}= \p _{k} \Gamma^{i}_{lj}-  \p _{l} \Gamma^{i}_{kj} +
 \Gamma^{i}_{ks} \Gamma^{s}_{lj}- \Gamma^{i}_{ls} \Gamma^{s}_{kj}.
\eeq
Fermion fields $\hi$ and $\lmi$ in (2.1) have integer spin zero and one
respectively. Integer spins of fermions arises naturally after twisting
\cite{Wsm} of the  N=2 \sm . On the other hand in the BRST approach
to TFT \cite{BS} fermions  $\hi$ and $\lmi$ arise as  ghosts and
therefore have integer spins. Their role is to cancel out boson degrees
of freedom in all correlation functions of observables. The field $\lmi$
satisfies the constraint
\beq
\lmi +\varepsilon^{\mu}_{\nu}J^{i}_{j}\lnj = 0.
\eeq

The model (2.1) is invariant under scalar fermionic symmetry generated by
the operator Q ($Q^{2}=0$), which can be considered as a BRST operator
\cite{BS,rev}. The transformation law for fields is
\beqa
\{Q,\wi\} & = & \hi,\\ \nonumber
\{Q,\hi\} & = & 0, \\ \nonumber
\{Q,\lmi \} & = & 2i(\p^{\mu}\wi - \varepsilon^{\mu\nu}J^{i}_{j}\dn\wj)
+\la^{\mu j}\Gamma^{i}_{jk}\hk.
\eeqa

Observables $O$ in TFT are elements of the Q-cohomology
\beqa
\{Q,O\} & = & 0, \\
\{Q,\tilde{O}\} & \neq & O.
\eeqa
Conditions (2.6),(2.7) mean that we should consider only gauge invariant
operators which are defined up to a gauge transformation.

Correlation functions we are interested in TFT are of the form
\beq
<O_{1}(x_{1}),\ldots O_{n}(x_{n})>, \label{cf for diff o}
\eeq
where $O_{i}$ are from Q-cohomology. Condition (2.7) means that if we
 substitute in (\ref{cf for diff o}) a Q-exact operator, then the correlation
function (\ref{cf for diff o}) would be zero, provided other $O_{i}$ satisfy
(2.6) and
\beq
Q\mid 0> = 0. \label{nonbr vac}
\eeq
 The latter condition holds true if the Q-symmetry is not broken.

Topological \sms (A-models, see the last ref. in \cite{CY/LG}) have two basic
properties which could be viewed as a definition of a TFT (cohomological one),
namely, energy-momentum tensor and the action are Q-exact
\beq
T_{\mu\nu}=\{Q,G_{\mu\nu}\}, \label{metric}
\eeq
\beq
S=\{Q,\Lambda\}. \label{coupling}
\eeq
In particular, ( \ref{metric}) means that correlation functions
(\ref{cf for diff o}) are world-sheet metric independent and (\ref{coupling})
means that they are coupling constant ($1/r^{2}$) independent. Indeed, the
variation of (\ref{cf for diff o}) with respect to the world-sheet metric
$h_{\mu\nu}$ results in the insertion of the Q-exact operator in
 (\ref{cf for diff o}) as it follows from (\ref{metric}),
 and therefore, gives zero.
In a similar way, the variation of  (\ref{cf for diff o}) with respect
to $r^{2}$ gives zero because of (\ref{coupling}).

The $h_{\mu\nu}$ independence of  (\ref{cf for diff o}) means, in particular,
its independence of the world-sheet coordinates $x_{i}$. The coupling
constant independence makes it possible to consider the theory in the
weak coupling limit $r^{2}\rightarrow \infty$. That means that the
semiclassical approximation gives actually  exact results. Thus
correlation functions  (\ref{cf for diff o}) are exactly given by their
instanton contributions.

For the $d=2$ model (2.1) there are only two local observables in the
cohomology (2.6),(2.7), namely
\beq
1 \mbox{ , } A_{ij}\hi \hj (x),
\eeq
where $A_{ij}$ is the antisymmetric tensor $dA=0$, $A\neq dB$. Tensor
$A_{ij}$ should have proper transformation law with respect to indices $i,j$
to preserve the invariance under target space diffeomorphisms. Thus the
natural choice for the only non-trivial local observable is \cite{Wgr}
\beq
O=i\J (w)\hi\hj (x). \label{O ind}
\eeq

Let us now calculate the instanton contribution to the correlation function
\beq
<O(x_{1}),\ldots O(x_{n})>, \label{cf}
\eeq
where the operator $O(x)$ is given by (\ref{O ind}).
We assume  that the target space metric $ \g$ and the complex structure
$\J$ have the form
\beqa
\g & = & \delta_{ij}g(w), \\ \nonumber
\J & = & \e_{ij}g(w).
\eeqa
Then the action (2.1) takes the form
\beqa
S & = & \frac{2r^{2}}{\pi}\intx
\left\{
g(w)\bp w \p \bw -\frac{i}{4}g(w)
\left[
\lap D\bar{\chi}
\right.
\right. \\ \nonumber
 & + &
\left.
\left.
\lam \bar{D}\chi
\right]
+ O(\lap \lam \bar{\chi}\chi)
\right\} \label{act}
\eeqa
Eq.(2.16) is written down in complex coordinates both for the world-sheet
and for the target space, for example, $z=x_{1}+ix_{2}$,
 $\p =1/2(\p_{1}-i\p_{2})$
and $w=w^{1}+iw^{2}$, $ \bw =\bw^{1}-i\bw^{2}$, $\lap = \la^{11} +i\la^{12}
+i\la^{21}-\la^{22}$.

Note that the boson term in the action (2.16) comes as a sum of the
kinetic term (the first term in (2.1))
\beq
S_{kin}=\frac{r^{2}}{\pi}\intx g(w)
\left(
\bp w \p \bw + \bp \bw \p w
\right) \label{kin}
\eeq
and the topological one (the second term in (2.1))
\beq
S_{top}=\frac{r^{2}}{\pi}\intx g(w)
\left(
\bp w \p \bw - \bp \bw \p w
\right) \label{top}
\eeq
The important feature of the \tsm (2.16) is that the
 topological term (\ref{top}) appears
with real coefficient in the action (the topological term with imaginary
coefficient could be added to the action as well), that means that
$\Theta$-angle is imaginary (see \cite{WYM} for the discussion of unitarity
and CPT-invariance).

The topological term (\ref{top}) is proportional to the winding number
 of the map $w(x)$ from the world sheet (which can be considered as $S_{2}$
with infinitely large radius) to the target space and thus is invariant
under continuous deformations of the function $w(x)$. Hence the functional
space is divided into topological classes with a given winding number $k$.

The minimum points of the action within each topological class are holomorphic
maps, which obey $\bp w=0$, namely
\beq
w(z)=v+ \sum_{l=1}^{k}\frac{\r_{l}}{z-z_{l}}, \label{mi}
\eeq
where $v$,$\r_{1}\ldots \r_{k}$ and $z_{1}\ldots z_{k}$ are $2k+1$ complex
parameters, while $k$ is the winding number of the instanton (\ref{mi}).
Note, that instantons with singular behavior at infinity can by obtained from
the one in (\ref{mi}) by coordinate inversion.

The action on the instanton (I) (\ref{mi}) is zero
\beq
S_{I}=0 \label{zero}
\eeq
because the contribution of the kinetic term (\ref{kin}) is exactly canceled
by the one coming from the topological term (\ref{top}).

 Let us now restrict ourselves to the simplest case of the I with the
winding number $k=1$
\beq
w_{I}(z)=v+ \frac{\r }{z-z_{0}}, \label{I}
\eeq
Here $x_{0}$ is the center of the I, $\mr$ and $arg(\r )$ are its size
and orientation.

I (\ref{I}) has six real (three complex) boson zero modes ($2(2k+1)$ in the
general case )
\beqa
\nu_{0} & = & \frac{\p w_{I}}{\p v}=1, \\ \nonumber
\nu_{1} & = & \r \frac{\p w_{I}}{\p \r }=\frac{\r }{z-z_{0}}, \\ \nonumber
\nu_{2} & = & \r \frac{\p w_{I}}{\p z_{0}}=\frac{\r^{2} }{(z-z_{0})^{2}}.
\label{bzm}
\eeqa
They satisfy the equation
\beq
\bar{D}\nu =0 \label{bzm eq}
\eeq
Consider now  fermionic zero modes of the I. They satisfy the equation
\beq
\bar{D}\chi =0,
\eeq
which is the same as in (\ref{bzm eq}). Thus we have six fermionic zero modes
($2(2k+1)$ in the general case)
\beqa
\chi_{0} & = & \al _{0}, \\ \nonumber
\chi_{1} & = & \frac{\al _{1}}{z-z_{0}}, \\ \nonumber
\chi_{2} & = & \frac{\al _{2}\r }{(z-z_{0})^{2}}.\label{fzm}
\eeqa
Here $\al _{0},\al _{1},\al _{2}$ are Grassmanian variables with
 dimensions (0, -1, -1).

Let us now calculate  correlation functions (\ref{cf}), where the
operator $O$ from  (\ref{O ind}) becomes
\beq
O=g(w)\bar{\chi}\chi . \label{O}
\eeq
The instanton measure has the  form
\beq
d\mu_{I}=-d^{2}vd^{2}x_{0}d^{2}\r d^{2}\al_{0} d^{2}\al_{1} d^{2}\al_{2}
 \label{imeasure}
\eeq
Note, that the factor $\exp{-const\;r^{2}}$ usual in instanton calculus
is absent in (\ref{imeasure}) due to the cancellation of the instanton action
in (\ref{zero}). This is in  accordance with the general rule which
requires the $r^{2}$-independence of (\ref{cf}) (more precisely, if we
introduce a new coefficient $\gamma$ different from $r^{2}$ in front
of the topological term in the action (2.1), then (\ref{cf}) could
depend only on the difference $r^{2}-\gamma$  [4] ). Note also
the cancellation of the determinants coming from boson and fermion
degrees of freedom in (\ref{imeasure}).

Now counting the number of fermion zero modes we conclude that only the three
point correlation function in (\ref{cf}) is nonzero (in general case
$n=2k+1$). This is a consequence of the $U(1)$ charge anomaly relation.
We have
\beqa
<O(x_{1}),O(x_{2}),O(x_{3})> & = & -\int d^{2}vd^{2}x_{0}d^{2}\r
 d^{2}\al_{0} d^{2}\al_{1} d^{2}\al_{2} \\ \nonumber
 &  & g(w)\bar{\chi}\chi (x_{1})
g(w)\bar{\chi}\chi (x_{2})g(w)\bar{\chi}\chi (x_{3})
\label{2.26}
\eeqa
Here fields $\chi ,\;\bar{\chi}$ should be understood as a sum over zero
modes (2.25). Now instead of integration over $v$, $\r$, $x_{0}$
in (2.28) let us proceed to new variables defined as follows.
Fix points $x_{1}$, $x_{2}$, $x_{3}$, and consider $w(x_{1})$, $w(x_{2})$,
$w(x_{3})$, as functions of $v$, $\r$, $x_{0}$ (see (\ref{I})). Then
using equations (2.25), (2.22) it is easy to see that functions
$\chi$ and $\bar{\chi}$ in (\ref{2.26}) represent the jacobian needed
to pass from $v$, $\r$, $x_{0}$ to $w_{1}=w(x_{1})$, $w_{2}=w(x_{2})$,
$w_{3}=w(x_{3})$. We get finally
\beq
<O(x_{1}),O(x_{2}),O(x_{3})>= \int d^{2}w_{1}g(w_{1}) d^{2}w_{2}g(w_{2})
d^{2}w_{3}g(w_{3}). \label{3p}
\eeq

The coordinate and coupling constant independence of the correlation
function becomes evident in (\ref{3p}) (provided the integrals are
convergent). The moduli space of instanton is just the product of
three copies of the target manifold and the correlation function
(\ref{3p}) is equal to the $A^{3}$, where $A$ is the area of the target
space. In general if $k\neq 1$ the correlation function (\ref{cf})
for $n=2k+1$ is given by the equation similar to the (\ref{3p}) one,
the moduli space of I being the product of $2k+1$ copies of the
target manifold.

In the next two sections we re-examine the above outlined procedure for the
calculation of correlation functions of observables (\ref{cf}) for
the special case of the \sigtag .

 \section{Cigar-like target space.}
\setcounter{equation}{0}

Let us consider the target space metric
\beq
g(w)=\frac{1}{1+ \mw ^{2}}. \label{bhm}
\eeq
Its difference from, say, the metric of the sphere for the $O(3)$ \sm
\beq
g(w)^{sphere}=\frac{1}{(1+ \mw ^{2})^{2}} \label{sphm}
\eeq
is in its slow fall-off  at large $\mw \rightarrow \infty$.

To see that (\ref{bhm}) represents the \bhm  let us perform the
change of variables
\beqa
w & = & \sinh{r}e^{-i\phi}\\ \nonumber
\bw & = & \sinh{r}e^{i\phi} \label{3.3}
\eeqa
Under this transformation the kinetic term for bosons (\ref{kin}) becomes
\beq
S_{kin}=\frac{r^{2}}{2\pi}\intx
\left\{
(\dm r)^{2}+ \tanh{r}^{2}(\dm \phi )^{2}
\right\}.
\eeq
The target manifold defined in (3.4) has the form of the semi-infinite
cigar. Its metric has been studied in \cite{Wbh}. In fact in ref.\cite{Wbh}
the gauged coset WZW model $SL(2,R)/U(1)$ has been considered. The
integration over gauge field leads to the theory of two real fields
$r$ and $\phi$  with the target space geometry given by (3.4). It has been
interpreted as a Euclidean version of the string theory for the two
dimensional black hole \cite{Wbh}. Note, that the black hole model
in \cite{Wbh} contains also a dilaton term added to the one in (3.4) which
is responsible for  its conformal invariance. The model we consider here
(defined by (2.16), (3.1)) has no dilaton term and therefore
is not a topological version of the  SL(2,R)/U(1) coset (the latter
is the special
case of the Kazama-Suzuki model \cite{KS}, see also \cite{MV}).

Let us now address a question \cite{Ybh}: do holomorphic instantons
(\ref{mi}) still exist in the \sm with metric (3.1). The main problem
is that the topological term
\beq
S_{top}\sim \frac{r^{2}}{\pi}\int \frac{dw\wedge d\bw}{1+ \mw ^{2}}
\eeq
becomes logarithmically divergent at large $w$. However, the coefficient
in front of the logarithm does not depend on the regularization scheme
and is still proportional to the winding number $k$. Indeed, substituting
(\ref{mi}) into (\ref{top}) and using (3.1) for the metric we get \cite{Ybh}
\beq
S_{top}=-2r^{2}k\left[log\frac{1}{a}+const \right],
\eeq
where $1/a$ is the UV cutoff. Thus, we still have the configuration space
divided into topological classes and  holomorphic instantons (\ref{mi})
are still minimum points of the kinetic term (\ref{kin}) in a given
topological class.

Above arguments led us to the conclusion that I (\ref{mi}) is a solution to
 equations of motion of the model (2.16), (3.1) and we can expand around
 them using the standard saddle-point method. Now let us check this directly.
Equations of motion for \sm (2.16) read
\beq
\frac{\delta S}{\delta \bw }= gD\bp w  =0
\eeq
However, as a matter of fact the r.h.s. is actually never zero on the
holomorphic instanton (\ref{mi}) even in \sm with a compact target space.
For example, for O(3) \sm with metric (3.2) one gets
\beq
\frac{\delta S}{\delta \bw }= gD\bp w_{I} \sim \sum_{l}(x-x_{l})^{3}
\delta^{(2)}
(x-x_{l}).
\eeq
To see what is going on let us expand the action around such a ``solution''
to equations of motion
\beqa
S(w_{I}+\nu ) & = & S(w_{I})+\intx \bar{\nu (x)}\frac{\delta S}
{\delta \bw _{I}} \\ \nonumber
 & + & c.c. + \frac{1}{2} \intx d^{2}y \nu^{i}(x)\frac{\delta^{2}S}{\delta
 w^{i}_{I}(x)\delta w^{j}_{I}(y)}\nu^{j}(y) + \cdots ,
\eeqa
where $w_{I}$ is the instanton solution and $\nu$ is a quantum fluctuation.
In order to use the standard saddle-point method what one really needs is
the absence  of the linear in $\nu$ term in (3.9). Substituting (3.8)
into (3.9) we indeed get zero linear term, provided $\nu (x)$ is non-singular
in the instanton center. \footnote{Quantum fluctuation $\nu$ is supposed to be
non-singular, otherwise it should be treated as another instanton, rather
 then the quantum fluctuation.} In this sense instanton is indeed a solution
to equations of motion in O(3) \sm .

Let us now turn to the \sm with \bhm (3.1). One has
\beq
\frac{\delta S}{\delta \bw }= gD\bp w_{I} \sim \sum_{l}(x-x_{l})\delta^{(2)}
(x-x_{l}).
\eeq
Plugging (3.10) into (3.9) one finds that the linear term is also zero in
this case. As we expected above the instanton (\ref{mi}) is a solution to
equations of motion for the \sm with \bhm  and one can expand around it
using the saddle-point method.

What about the instanton action, which is the sum of (\ref{kin}) and
 (\ref{top}). Note, that both the kinetic term (\ref{kin}) and the
topological one (\ref{top}) are logarithmically divergent on I.
Due to the singularities introduced by $\delta$-functions the instanton action
turns out to be nonzero in the contrast to the case of the compact target
space (\ref{zero}).\footnote{The author is indebted to A. Losev for this
remark.} Let us calculate it. One has
\beq
S_{I}=\frac{2r^{2}}{\pi}\intx \frac{\bp w_{I}\p \bw _{I}}
{1+\mid w_{I}\mid ^{2}}.
\eeq
Substituting $\bp w_{I}=\pi \sum_{l}\r _{l}\delta^{(2)}(x-x_{l})$ and
using some regularization for the $\delta$-function, say,
\beq
\delta^{(2)}(x)=\frac{1}{\pi}\frac{a^{2}}{(x^{2}+a^{2})^{2}}, \label{delta}
\eeq
one gets
\beq
S_{I}=k\frac{r^{2}}{3}. \label{r/3}
\eeq
We see that the logarithmic divergences are canceled out in (\ref{r/3}),
the instanton action is nonzero but finite. Eq. (\ref{r/3}) means that
instanton contribution to correlation functions are weighed with the
factor $g_{I}^{k}$, where
\footnote{Note that the exact coefficient in front of $r^{2}$ in
 (\ref{r/3}) and
 (\ref{gi}) could depend on the regularization scheme. Therefore, in
what follows we consider $g_{I}$ as a new independent coupling constant.
It turns out to be the true coupling constant of the model, since there is
no direct dependence of  observables on $r^{2}$.}
\beq
g_{I}=e^{-\frac{r^{2}}{3}}. \label{gi}
\eeq
The result  (\ref{r/3}) means that instantons are not suppressed in the
the \tsm with the \bhm (3.1).

We would like to stress now that the target space with
metric (3.1) is the limiting case of a
noncompact manifold for which unsuppressed instantons exist. Indeed, consider
the metric
\beq
g_{q}(w)=\frac{1}{(1+\mw ^{2})^{q}} \label{q-m}
\eeq
At $q>1$ the target space is compact and we have conventional instantons
with zero action in the topological version of the \sm . At $q=1$ we have
instanton solutions (as it follows from (3.10)) with finite action
 (\ref{r/3}). At $1/2<q<1$ we still have instanton solutions to equations
of motion as it follows from the considerations similar to the above ones
(see eqs. (3.8)-(3.10)). However, the instanton action diverges as
$1/a^{2(1-q)}$, hence contributions of these instantons are suppressed
in all the correlation functions. At $q<1/2$ we have no instantons as
 holomorphic curve  (\ref{mi}) fails to be a solution to equations of motion.

We see that $q=1$ case is on the borderline between the compact and noncompact
target space models. Moreover, the  \bhm , perhaps, should be thought of as
being ``compact''  because there are nonsuppressed instantons in the
topological version of \sm  with this metric.

The existence of unsuppressed instantons in the \tsm with the \bhm  is one
of the main results of the present paper. These instantons have unusual
physical properties. In particular, as we show below the moduli space
of these instantons is noncompact. Therefore, new logarithmic divergences
emerge in the theory. We study some of these effects in the remainder
of the paper.

\section{Correlation functions}
\setcounter{equation}{0}

Now let us calculate correlation functions  (\ref{cf}) of the observable
 (\ref{O}) for the \sm with \bhm (3.1).
Again for the sake of simplicity let us restrict ourselves to the case
of the instanton  (\ref{I}) with the winding number $k=1$. Zero modes of I is
given essentially by eqs. (2.22),  (2.25). The only
exception is that we are not going now to integrate over parameter $v$
in the instanton measure. The reason is that $\nu_{0}$, $\chi_{0}$
zero modes in  (2.22),  (2.25) have the quadratically divergent
norm on the world-sheet taken to be a complex plane. Moreover, $v$ is
a boundary condition for field $w$ at infinity. In fact it has a meaning
of vacuum expectation value (VEV) $<w>$.

Usually we do not integrate over VEV in QFT. That would mean summing
up all the different vacuums of the theory. Instead, we minimize the
vacuum energy with respect to VEV to find the true vacuum. Of course, if
physics do not depend on VEV (as is the case, say,  for the phase of the scalar
field in the Gauge-Higgs model
 )  then one can safely integrate over it; nothing would change except
some numerical factor in front of the partition function.
However, once we suspect that physics could be $\mv $ dependent, in the
case at hand, it is better to keep $v$ fixed.

Thus the instanton measure now is
\beq
d\mu_{I}=-g_{I}d^{2}x_{0}d^{2}\r d^{2}\al _{1} d^{2}\al _{2},\label{im}
\eeq
instead of (\ref{imeasure}). Here $g_{I}$ is given by  (\ref{gi}).
Eq.  (\ref{im}) leads to another selection rule for correlation
functions, namely  (\ref{cf}) is nonzero if $n=2k$. Thus for $k=1$
we have $n=2$. Repeating the same calculation, which leads us from
(2.28) to (\ref{3p}) one gets in the case at hand
\beq
<O(x_{1}),O(x_{2})>= -\int d^{2}w_{1}g(w_{1}) d^{2}w_{2}g(w_{2})
. \label{2p}
\eeq
Note that deriving this equation we have not actually used the concrete
form of the \bhm (3.1). Therefore, (\ref{2p}) holds true also for
\tsm with compact target space (with coupling constant $g_{I}=1$). We see,
that our prescription not to integrate over constant modes defines
another formulation of the theory based on the same Lagrangian
 (2.16). It is of course
also a topological one (for the \sm with compact target space  )as it is clear
already from  (\ref{2p}). The new selection rule $n=2k$ means, in particular
that the partition function $Z=const$ instead of being zero (which is
more usual in conventional QFT), while
$<O(x)>=0$. In what follows we study \tsm in this formulation. The physical
motivation for this as discussed above,  is based essentially on the
 observation that it is not very useful to integrate over VEV of the field $w$
if physics depend on it.

Now let us specialize for the case of the \bhm (3.1). Substituting (3.1)
into  (\ref{2p}) we see that integral over moduli space of I becomes
logarithmically divergent ( like  ones for topological and kinetic terms,
see (3.6)) and needs regularization. We regularize the integral in
 (\ref{2p}) in the same way as before, introducing UV and IR cutoffs
on the world-sheet, $1/a$ and $1/L$ respectively. Roughly speaking,
  (\ref{2p}) gives the answer $\sim log^{2}\mid w_{max}\mid $. To see what
is $\mid w_{max}\mid$ in terms of cutoff parameters on the world-sheet
one recalls that
\beqa
w_{1} & = & v+\frac{\r}{z_{1}-z_{0}}, \\ \nonumber
w_{2} & = & v+\frac{\r}{z_{2}-z_{0}}. \label{4.3}
\eeqa
Taking into account  (4.3) one gets with the double logarithm
accuracy at finite $\mv$ ($\mv \ll L/a$)
\beq
<O(x_{1})O(x_{2})>=-2(2\pi )^{2}g_{I}\log{\frac{\mid x_{12} \mid}
{a}}\log{\frac{L}{\mid x_{12} \mid}}+ O(log), \label{anw}
\eeq
where $x_{12}=x_{1}-x_{2}$. The $ x_{12}$-dependence here comes as
follows. To get the double logarithm one has to consider either
region $x_{0}\rightarrow x_{1}$ ($a\ll \mid x_{0}-x_{1}\mid \ll \mid x_{12}
\mid $ or
region $x_{0}\rightarrow x_{2}$ ($a\ll \mid x_{0}-x_{2}\mid \ll \mid x_{12}
\mid $ ). Suppose  $x_{0}\rightarrow x_{1}$, then the integral over $\mid w_{1}
\mid $ goes from $\r /\mid x_{12} \mid $ to $\r /a$ and gives $\log{\mid x_{12}
\mid /a}$. Further, at  $x_{0}\rightarrow x_{1}$ $\mid w_{2\;max} \mid =L/
\mid x_{12}\mid $. This gives $\log{\frac{L}{\mid x_{12} \mid}}$. Factor 2
arises when one takes onto account the second option  $x_{0}\rightarrow x_{2}$.
These arguments are just to explain qualitatively the origin of the answer
(\ref{anw}). Of course, the integral in (\ref{2p}) can be calculated in a
more rigorous way.

For the sake of a future use let us calculate (\ref{2p}) also in the limit
of $\mv \rightarrow \infty$, namely $\mv \gg L/a$. Then $\mid w_{1} \mid
\sim \mid w_{2} \mid \sim \mv $ and one gets
\beq
<O(x_{1})O(x_{2})>=-\pi^{2}\frac{L^{4}}{\mid x_{12}\mid ^{2}a^{2}}
\frac{1}{\mv ^{4}}\rightarrow 0. \label{4.5}
\eeq
This goes to zero as $\mv \rightarrow \infty$.

The correlation function in (\ref{anw}) shows $x_{12}$-dependence. This is a
signal for the  BRST-symmetry breaking. Indeed, the derivative of an
element of the Q-cohomology is always Q-exact \cite{WYM}. Explicitly
\beq
\dm O(x)=\{Q,iJ_{ij}\dm w^{i}\chi^{j}\} .
\eeq
Hence, nonzero value for $<\dm O(x_{1}O(x_{2})>$ means that
\beq
Q\mid 0>\neq 0,
\eeq
and the BRST-symmetry is broken.

Now let us observe that we run into a problem. Indeed, the \tsm with d=2
has the value of the Witten index \cite{Wind} $I_{W}=2 \neq 0$ (it is
equal to the number of the physical states, which is the same as the number of
elements of the Q-cohomology, see for example \cite{rev}). Hence, supersymmetry
cannot be broken in the model at hand. So, what is going on?

One suggestion which can save the day is that the BRST-symmetry is in fact
not broken for the  true stable vacuum state of the theory which corresponds
to $\mv \rightarrow \infty$. Indeed,  (\ref{4.5}) shows that at
 $\mv \rightarrow \infty$ $<O(x_{1})O(x_{2})> \rightarrow 0$ and thus shows
no $x_{12}$-dependence in this limit.

We show in the remainder of this paper that this assumption is correct.
In fact, in Sect. 6 we show that for finite $\mv$
the model possess an infinite number
of degenerative vacuum states with nonzero vacuum energy $E_{vac}$. However,
as $\mv\rightarrow \infty$ $E_{vac}\rightarrow 0$. Thus, the theory has a
stable vacuum state at $\mv \rightarrow \infty$ for which BRST-symmetry
is not broken. This resolves the puzzle with the nonzero value
of the Witten index. However, for the unstable vacuum states at finite
$\mv$ we have $x_{12}$-dependent correlation function  (\ref{anw})
and the BRST-symmetry broken down.

The similar phenomenon of ``vacuum running to infinity''
is known to occur in the 4D SUSY QCD \cite{Aff}.
Instantons generate the superpotential which removes the
degeneracy of the vacuum states associated with different
values of the VEV of the scalar field. As a result the true vacuum
state with zero vacuum energy corresponds only to VEV going to infinity.
The theory also has nonzero Witten index.

\section{Anti-instantons}
\setcounter{equation}{0}

Our aim now is to show that the vacuum energy in the \sm with \bhm  is nonzero
for finite $\mv$. However, it is clear that the perturbative theory gives
zero for $E_{vac}$ since our model is a topological one at the
perturbative level. Instantons by themselves do not contribute to the $E_{vac}$
because of the anomalous selection rule n=2k (instantons have fermion zero
modes). What we need to get the nonzero vacuum energy are the
instanton-anti-instanton pairs (\II). In order to calculate the contribution
of \II  into $E_{vac}$ we study anti-instantons (\bi ) in the model at hand
in this section and derive the effective Lagrangian which takes them
into account.

\bi  is a anti-holomorphic map from the world-sheet to the target
\beq
w=v+\sum_{l=1}^{p}\frac{\br _{al}}{\bar{z}-\bar{z_{al}}}.
\eeq
Here $-p$ is the winding number, $x_{al}$, $\br_{al}$ are new complex
parameters. Of course, (5.1) satisfies equations of motion in the
same sense as I (2.19) do (see (3.10))
\beq
\bar{D}\p w \sim 0
\eeq
Again we restrict ourselves to \bi  with p=1:
\beq
w=v+\frac{\br _{a}}{\bar{z}-\bar{z_{a}}}.
\eeq
where $x_{a}$ and $\br_{a}$ has the interpretation of the  position and the
orientation-size vectors.

In nontopological versions of \sms I's and \bi's come on the  same ground.
Instead, in \tsms with compact target space I's come with the zero
action (\ref{zero}) while
\beq
S_{\bar{I}}=2r^{2}A.
\eeq
Indeed, the topological term (\ref{top}) cancels the contribution of
the kinetic one in (2.16) for I and doubles it for \bi . Now it
is clear  that \bi 's cannot contribute to
correlation functions in the \tsm with unbroken BRST-symmetry. This is
a consequence of the general statement of $r^{2}$-independence of
observables. Indeed, \bi  contribution would involve the factor
$\exp{-pS_{\bar{I}}}$ in the  contradiction with the general rule.

Now we are going to study \bi 's in the \tsm with metric (3.1) and show
that they do contribute to observables of the model at finite $\mv$.

Zero modes of \bi  (5.3) can be studied in a way similar to that for I
 (\ref{I}). We have four boson zero modes which are associated with variations
of (5.3) with respect to $x_{a}$ and $\r_{a}$ like in  (2.22) ( note,
that we keep $v$ fixed). Furthermore, we have four fermion zero modes. They
are
\beqa
\lambda^{++}_{1} & = & \frac{\bt_{1}}{\bar{z}-\bar{z_{a}}}, \\ \nonumber
\lambda^{++}_{2} & = & \frac{\br_{a}\bt_{2}}{(\bar{z}-\bar{z_{a}})^{2}}.
\eeqa
They satisfy the equation
\beq
D_{\mu}\lmi =0,
\eeq
which reduces to $\p \lambda^{++}=0$.

 The action on \bi  (5.3) is logarithmically divergent as it is clear from the
discussion above. It reads
\beq
S_{\bar{I}}=4r^{2}\log{\frac{\mid \r_{a}\mid}{a}}.
\eeq
Eq. (5.7) means that large size \bi 's are suppressed in the path integral.
However,  small size \bi 's (with size $\mid \r_{a}\mid \sim a$) can induce
a new point-like interaction (see \cite{Ybh}).

In general, the effect of instantons in any theory in which they are present
can be described by means of the effective Lagrangian \cite{CDG,Ybh}. In
$d=2$ \sm this \bi -induced vertex is (cf. \cite{Ybh})
\beqa
V_{\bar{I}} & = & -\int d\mu_{\bar{I}}e^{-S_{\bar{I}}}
\left(
\frac{ir^{2}}{2}
\right)^{4}
g^{4}(v)\bar{\bt_{1}}\chi \bt_{1}\bar{\chi} \\ \nonumber
 &  & \r_{a}\bar{\bt_{2}}\p \chi \bar{\r_{a}}\bt_{2}\bp \bar{\chi}
exp \{ 2r^{2}g(v)[\r \p w +\br \bp \bw ] \},
\eeqa
where $d\mu_{\bar{I}}$ is the \bi  measure to be discussed below.
This vertex should be added to the action (2.16) to get the
effective action which mimics the effects of \bi 's at the
perturbative level. To check this let us calculate the following
correlation function
\beq
<w(x_{1})\ldots w(x_{n})\lambda^{++}(x)\lambda^{++}(x')>_{\bar{I}}
\eeq
in the one-\bi  background.

On the one hand (5.9) can be calculated (in the leading order in $1/r^{2}$)
substituting classical expressions (5.3) and (5.5) for fields $w$ and $\lambda$
 into (5.9). The result is
\beq
\prod_{i=1}^{n}\left(v+\frac{\bar{\r_{a}}}{\bar{z_{i}}-\bar{z_{a}}}\right)
\frac{\bt_{1}}{\bar{z}-\bar{z_{a}}}\frac{\bar{\r_{a}}\bt_{2}}
{(\bar{z'}-\bar{z_{a}})^{2}}.
\eeq
On the other hand the same result can be reproduced in the purely
perturbative manner, inserting (5.8) into the action (2.16) of the
model. Indeed, taking in the expansion of $\exp{-V_{\bar{I}}}$ the only
first power of $V_{\bar{I}}$ ( this corresponds to the one-\bi  contribution)
 one gets the same answer as in (5.10) contracting fields in
(5.9) with fields in (5.8) and taking into account the propagation functions
of the model
\beqa
<w(x),\bw (0)> & = & \frac{1}{g(v)r^{2}}\log{\frac{L}{\mid x\mid}} +v^{2},
\\ \nonumber
<\lambda^{++}(x),\bar{\chi}(o)> & = & -\frac{2i}{g(v)r^{2}} \frac{1}{\bar{z}},
\\ \nonumber
<\lambda^{--}(x),\chi(o)> & = & -\frac{2i}{g(v)r^{2}} \frac{1}{z}.
\eeqa
Note that $v$ is the VEV of the field $w$ one expands around in the
perturbation theory.

 We have derived the vertex (5.8) above only in the semiclassical
approximation $r^{2}\gg 1$. To get also quantum corrections one can use
the following trick: substitute $w$ instead of $v$ in factors $g(v)$
which appear in (5.8). It is clear that the effective Lagrangian should
depend on the field $w$ rather then its VEV if there is no explicit
symmetry breaking. One gets
\beqa
V_{\bar{I}} & = & -\frac{r^{8}}{16}\int d\mu_{\bar{I}}e^{-S_{\bar{I}}}
\mid \r_{a}\mid^{2}
g^{2}(w)\bar{\bt_{1}} \bt_{1}\bar{\bt_{2}} \bt_{2}
g(w)\bar{\chi}\chi \\ \nonumber
 &  & g(w) \bar{D} \bar{\chi}D\chi
exp \{ 2r^{2}g(v)[\r \p w +\br \bp \bw ] + O(r^{2}\r^{2}\p^{2}w)\}
\eeqa
 Here $ O(r^{2}\r^{2}\p^{2}w)$ denotes the possible contributions of higher
 derivative terms $\r^{n}\p^{n}$ which is beyond  our control here.
Note, that the trick of the substitution  $v\rightarrow w$ gives us nontrivial
quantum corrections in (5.12). To check these explicitly one has to go through
two-loop (and higher loop) calculation in the \bi   background.\footnote{Cf.
\cite{Ybh} where these corrections remain  undetermined because the
VEV v of the field w have been set to zero} Note, also the obvious
generalization $\p \chi \rightarrow D\chi$, $\bp \bar{\chi}\rightarrow
\bar{D}\bar{\chi}$ which also accounts for quantum corrections.

Now let us discuss what $\bar{I}$ measure $d\mu_{\bar{I}}$ in (5.12) is.
The following part of this section is rather technical, therefore the reader
who is more interested in physics than in technical details could go straight
to our final result for $V_{\bar{I}}$ in eqs.(5.26), (5.28).
According to general rules the \bi  measure yields (we assume $r^{2}>>1$) here
to justify the one loop calculation.
\beq
d\mu_{\bar{I}}=\frac{d^{2}x_{a}d^{2}\r _{a}}{a^{4}}d^{2}\beta _{1}d^{2}\beta
_{2}\left(\frac{det^{'}\Box ^{\bar{I}}_{B}}{det\Box_{B}^{0}}\right)^{-1/2}
\left(\frac{det^{'}\Box_{F}^{\bar{I}}}{det\Box_{F}^{0}}\right)^{1/2}
\eeq
Here $\Box_{B}^{\bar{I}}$ is the operator of the quadratic fluctuations in the
$\bar{I}$ background
\beq
\Box^{i}_{Bj}=-[D^{i}_{\mu p}D^{p}_{\mu j}+R^{i}_{l,jk}(w)\dm w^{k}
\dm w^{l}],
\eeq
where $R^{i}_{l,jk}$ is the curvature tensor for the target space (2.3)
while $\Box^{0}_{B}$ is the same operator for the trivial background field
 $w=v$ .
In an analogous way $\Box_{F}^{\bar{I}}$ is the fermion operator
\beq
\Box^{i}_{Fj}=-[D^{i}_{\mu p}D^{p}_{\mu j}-\frac{1}{2}\emn J^{i}_{k}
R^{k}_{j,nm}(w)\dm w^{n}\dn w^{m}]
\eeq
Prime over det's in (5.13) means the product of nonzero eigenvalues. It is a
rather complicated problem to compute determinants in (5.13) exactly. However,
 we need not do that. All we need is the dependence of $d\mu_{\bar{I}}$
 on $\r _{a}$ and $v$ , which fix $d\mu _{\bar{I}}$ up to a numerical constant.
The dependence on $\r _{a}$ is easy to calculate since $\r _{a}$ can enter
determinants in (5.13) only as a ratio $\mid \r_{a}\mid /a$ .The latter is
completely
fixed by the one loop metric-complex structure $\beta $ -functions (note that
the model at hand is not a conformal one).

So let us first discuss $\beta $-functions. As is is well known the first
coefficient
of the metric $\beta $-function is proportional to the Ricci tensor of the
target
space \cite{MS}. Explicitly, expanding det $\Box_{B}$ in powers of any given
external field $w$ and taking into account the term with $log 1/a$ one gets
\beqa
 & - & \frac{1}{2}\log{\left(\frac{det'\Box_{B}}{det\Box_{B}^{0}}\right)}
\mid _{log
1/a}  =  \log{\frac{\mid \r_{a}\mid}{a}}
\left\{\frac{1}{4\pi}\intx R_{ij}\dm w^{i}\dn w^{J} -4 \right\}
\\ \nonumber
 & = &  \log{\frac{\mid \r_{a}\mid}{a}}
\left\{-\frac{1}{\pi}\intx (\p w \bp \bw + \p \bw \bp w)
\left[\frac{g'}{g}+\left(\frac{g''}{g}-\frac{g'^{2}}{g^{2}} \right)
\mid w \mid ^{2} \right] -4 \right\}.
\eeqa
Here $R_{ij}=R^{k}_{i,kj}$ and $g$ is considered  as a function of $\mid w
\mid ^
{2}$ ,  $g'$  is its derivative with respect to $\mid w\mid ^{2}$. Term $
-4log\mid \r_{a}\mid /a$ accounts for the subtraction of zero modes.
It just cancels the factor $1/a^{4}$ in (5.13) as the a-dependence of
$d\mu_{\bar{I}}$ can come only from coupling constant renormalization (and
 from a-dependence of $S_{\bar{I}}$ in (5.7) for the particular case of the
black hole metric). The size of $\bar{I}\mid \r_{a}\mid $ is put in (5.16)
 just as a IR cutoff.
In a similar way one can calculate the fermion contribution to one-loop
 $\beta$-function\cite{Horn}. One gets the complex structure renormalization
\beqa
&  & \frac{1}{2} \log{\left(\frac{det'\Box_{F}}{det\Box_{F}^{0}}\right)}
\mid _{log1/a}  =  \log{\frac{\mid \r_{a}\mid}{a}}
 \left\{\frac{1}{8\pi}\intx
 \emn J^{i}_{k}R_{i,nm}^{k}\dm w^{n}\dn w^{m}\right\}
 \nonumber \\
 & = &  \log{\frac{\mid \r_{a}\mid}{a}}  \nonumber \\
 & &
\left\{\frac{1}{\pi}\intx (\p w \bp \bw - \p \bw \bp w)
 \left[\frac{g'}{g}
 +  \left(\frac{g''}{g}-\frac{g'^{2}}{g^{2}} \right)
\mid w \mid^{2} \right]\right\}.
\eeqa
We observe now that for $I$ background $(\bp w=0)$ both factors (5.16) and
 (5.17)
 cancels each other. This, of course, should be expected: as there is no
coupling constant dependence (more generally, no target space metric $g$
dependence)
of $I$ contributions at the classical level, coupling constant (more generally
metric ) renormalization can not arise as well. Instead, for $\bar{I}$
background  $(\p w=0)$ the fermion $\beta $ -function in (5.17) doubles the
boson one in
(5.16). Thus the measure $d\mu_{\bar{I}}$ is proportional to
\beq
\left(\frac{\mid \r_{a}\mid}{a}\right)^{2b},
\eeq
where b is
\beq
b=-\frac{1}{\pi}\int d^{2}x\p \bw _{\bar{I}}\bp w _{\bar{I}}\left[
\frac{g'}{g}+\left(\frac{g''}{g}-\frac{g'^{2}}{g^{2}}\right)\mid w_{\bar{I}}
\mid ^{2}\right]
\eeq
Say, for O(3) $\sigma $-model  $b^{sphere}=2 $ , moreover, renormalization (5.
16), (5.17) does not change the form (3.2) of the metric and complex structure.
 For black hole metric (3.1)
\beq
b^{BH}=1,
\eeq
however, the form of the metric and complex structure in (3.1) is not
stable under renormalization (5.16),(5.17).

Now let us discuss the $g(v)$ dependence of the measure in (5.10).
It is very important for us because we are going to study the $\mv $
 dependence of $V_{\bar{I}}$ .To calculate it we use the method of ref.
\cite{FFS}
 which allows us to compute the variation of determinants in (5.13) with
respect to given parameters (now we are interested in variations with respect
 to $v$ ) in terms of low eigenvalue and high eigenvalue limits of certain
 exponential operators. One has, say, for $det\Box_{B}$
\beqa
-\frac{1}{2}\delta_{v}\left(\log{\frac{det'\Box_{B}}{det\Box_{B}^{0}}}
\right) & = & -\frac{1}{2}Tr\left\{g^{-1}\delta _{v}g\left(e^{-t\Box _{B}}-
e^{-t\bigtriangleup_{B}}\right)\right\}\mid _{t=\infty}\\ \nonumber
 & + & \frac{1}{2}Tr\left\{
g^{-1}\delta _{v}g\left(e^{-t\Box_{B}}-e^{-t\bigtriangleup_{B}}\right)\right\}
\mid _{t=0}.
\eeqa
Here the trace is understood as both the trace of operator and the one over
target space
 induces, while
\beq
\bigtriangleup^{i}_{Bj}=-\left(D^{i}_{\mu p}D^{p}_{\mu j}-
R^{i}_{l,jk}\dm w^{k}\dm w^{l}\right).
\eeq
In fact one should be more careful considering the high eigenvalue term
$(t=0)$ in (5.21) \cite{FFS}. Namely, one should introduce the world-sheet
 metric
$h_{\mu \nu }(x)$ because this term gives
\beq
\frac{d}{8\pi }\int d^{2}x\delta_{v}[\log{g}]R^{(2)},
\eeq
where $R^{(2)}$ is the world sheet curvature and $d=2$ .This is nothing
other than the dilaton $\beta $ -function related to the Polyakov anomaly
\cite{P}.\footnote{ It is curious to note, that this term had been
calculated in \cite{FFS} before the Polyakov's original paper \cite{P}
appeared however, was not interpreted in terms of anomaly.}
The high eigenvalue term of the fermion determinant gives the same result as
in (5.23) but with the inverse sign, thus both factors cancels each other.
This should be expected since the total central charge is zero in the TFT.

The low eigenvalue term in (5.21) is the contribution of zero modes.
It gives $\delta_{v}(S_{\bar{I}}g(v)\log{L/a})^{-1}$.
For the $\bar{I}$ background the fermion low eigenvalue term gives the same
result, thus one gets the factor
\beq
\frac{1}{g^{2}(v)S^{2}_{\bar{I}}log^{2}L/a}
\eeq
Now, taking into account (5.18) and (5.24) one obtain
\beq
d\mu _{\bar{I}}=const\frac{d^{2}x_{a}d^{2}\r _{a}}{\mid \r _{a}\mid ^{4}
g^{2}(v)}\frac{1}{S^{2}_{\bar{I}}}d^{2}\beta _{1}d^{2}\beta _{2}
\left(\frac{\mid \r _{a}\mid }{a}\right)^{2b}
\eeq
Making the substitution $v\rightarrow w$ in the argument of metric here
 and inserting the result in (5.12) one gets finally (after integration over
$\beta _{1}$,$\beta _{2}$)
\beqa
V_{\bar{I}} & = & -g_{\bar{I}}\int d^{2}x
\frac{d^{2}\r _{a}}{\mid \r _{a}\mid ^{2}}
\frac{1}{S_{I}^{2}}\left(\frac{\mid \r _{a}\mid}{a}\right)^{2b}
e^{-S_{\bar{I}}}g(w)\bar{\chi}\chi
\dm ^{2}\left(g(w)\bar{\chi}\chi\right) \\ \nonumber
 & & exp\left\{2r^{2}g(w)\left[
\r _{a}\p w+\bar{\r _{a}}\bp \bw \right]+O(r^{2}\r _{a}^{2}\p ^{2}w)\right\},
\eeqa
where we introduced a constant $g_{\bar{I}}$ ,to be treated as a new
 independent coupling constant of the model. Operators $\chi $, $\bar{\chi}$
here could give nonzero contributions only if we consider them in the
external instanton field. Therefore, one can replace
$g\bar{D}\bar{\chi}D\chi$ in (5.12) by $\dm ^{2}\left(g\bar{\chi}\chi\right)$
using the fact that $D\bar{\chi} =0$, $\bar{D}\chi =0$ for I field.
The effective vertex (5.26) holds true for any $d=2$ topological $\sigma$-
model. The only factors which depend on the particular metric are $S_{I}$ and
b.

It is easy to see that $V_{\bar{I}}$ in (5.26) is a Q-invariant one.
Moreover, it is Q-exact. This can be checked using eq.(4.6), which shows
that operator $\dm ^{2}O(x)$ in (5.26) is Q-exact.
Therefore, indeed, if the BRST-symmetry is not broken then $\bar{I}'s $
plays no role, in accordance with the general rule.

Of course, the similar effective vertex can be obtained for instantons.
Let us write it down for the sake of completeness(we are not going to use it
in this paper).
\beqa
V_{\bar{I}} & = & -g_{I}\int d^{2}xd^{2}\r \mid \r \mid ^{2}g^{2}(w)g(w)
\lam\lap g(w)D\lam\bar{D}\lap
 \\ \nonumber
 & & exp\left\{2r^{2}g(w)\left[\r \p \bw +\bar{\r }\bp w\right]+
O(r^{2}\r ^{2}\p ^{2}w)\right\}.
\eeqa
Effective vertices (5.26) (5.27) reproduce effects of $I's$ and $\bar{I}'s$
in the framework of the perturbation theory. They have to be added to the
original action (2.16) of the model which afterwards should be treated
perturbatively. In particular, expanding $exp-V$ in powers of $V_{I}$, $V_
{\bar{I}}$ and taking into account nonlinear terms in
$V$ one can study effects of $I\bar{I}$ interactions in the instanton
medium\cite{CDG}.

Now let us concentrate on the case of the black hole metric (3.1). The
$\bar{I}$ action has the form (5.7). Therefore, the integral over
$\r _{a}$ in (5.26) is UV divergent at $r^{2}>>1$ and could be
performed in the closed form. Cutting it off at $\mid \r _{a}\mid \sim a $
one gets
\beq
V_{\bar{I}}=-g_{\bar{I}}\int d^{2}xg(w)\bar{\chi}\chi\dm ^{2}
\left(g(w)\bar{\chi}\chi\right).
\eeq
To conclude this section, let us observe that the vertex (5.28) is invariant
under global target space diffeomorphisms. Indeed, it can be rewritten as
\beq
V_{\bar{I}}=-g_{\bar{I}}\int d^{2}xiJ_{ij}(w)\chi ^{i}\chi ^{j}\dm ^{2}
\left(iJ_{kl}(w)\chi ^{k}\chi ^{l}\right),
\eeq
which is manifestly covariant. This provides a check for our computation of
the
$v$ dependence of the $\bar{I}$ measure in (5.26). The vertex in (5.28)
is also obviously Q-invariant as it is constructed with the help of operator
(2.13) which is the one from the Q-cohomology of the $\sigma $-model.

\section{Vacuum energy.}
\setcounter{equation}{0}

Now let us use the effective $\bar{I}$ vertex in (5.28) to calculate the
$I\bar{I}$ contribution to the vacuum energy. Observe first that this
calculation is essentially the same as the one for the correlation function
(4.2)
in the one $I$ background we already performed in section 4. Indeed, (5.28)
gives
\beq
E^{I\bar{I}}_{vac}=<V_{\bar{I}}>_{I}=-g_{\bar{I}}\p ^{2}_{x_{1}}<O(x_{1})
O(x_{2})>\mid _{x_{2}\rightarrow x_{1}}
\eeq
 The eq. (6.1) shows that if there were no $ x_{12} $ - dependence of the
correlation
 function (4.2) the vacuum energy would be zero. More generally speaking
 $V_{\bar{I}} $ in (5.28) is obviously Q-exact. Thus, if the BRST-symmetry
is not broken, then $ E_{vac}=0$ in (6.1) as should be the case on the
general ground. However, once we get $ x_{12} $-dependence of
$<O(x_{1})O(x_{2})>$ for finite $\mid v\mid $ in eq.(4.4), the vacuum energy
is nonzero. Indeed, plugging (4.4) into (6.1) one arrives at finite $v$ at
\beq
E_{vac}=16\pi^{2}g_{I}g_{\bar{I}}\frac{\left[\log{L/a}+O(1)\right]}
{a^{2}}V^{(2)},
\eeq
where $V^{(2)} $ is the volume of the world sheet. We use, here the
regularization of the $\delta $ function (3.12) which gives $\delta ^{(2)}
(0)=\frac{1}{\pi a^{2}}$. Instead, if $\mid v\mid \rightarrow\infty $,
 namely $\mid v\mid >>L/a $ one uses eq.(4.5) for the correlation function in
(6.1). One gets the vacuum energy
\beq
E_{vac}=\frac{4\pi ^{2}g_{I}g_{\bar{I}}}{a^{2}}\frac{L^{4}}
{\mid v\mid ^{4}a^{4}}V^{(2)},
\eeq
which goes to zero at $\mid v\mid \rightarrow\infty $.

 Results (6.2), (6.3) show that at finite $\mid v\mid $ we have an infinite
number of unstable degenerative vacuum states (each of which corresponds
to a given $\mid v\mid $ ) with nonzero vacuum energy (6.2). Hence the
 BRST-symmetry is broken for these states. Moreover, the correlation
function (4.4) shows the coordinate dependence. Thus, the theory at
finite $\mid v\mid $ can be interpreted as a broken phase of  the TFT.

Instead, for $\mid v \mid \rightarrow\infty $ we have $ E_{vac}
\rightarrow 0$. On the other hand the correlation function in (4.5)
looses its $x_{12}$-dependent correlation function
$<O(x_{1})O(x_{2})>$. The BRST symmetry is not broken in that stable
vacuum state in accordance with the nonzero value of the Witten index.
This state is interpreted as a topological phase of the model at hand.

\section{Discussion}
\setcounter{equation}{0}

In this paper we study the topological $\sigma $ model with the black
 hole target space metric (3.1). We show that there are peculiar
instantons in that model with the noncompact moduli space. As a result of
divergences of integrals over the moduli space of instantons the unbroken
topological phase is realized only at the VEV of the field $w$ going to
infinity. Instead, at finite $\mid v \mid$ the theory possess an infinite
number of unstable degenerative vacuum states with nonzero vacuum energy
(6.2) and x-dependent correlation function (4.4). These vacuum states
correspond to broken phase of the TFT.

The breakdown of the BRST-symmetry for these states can be considered
as a spontaneous one. Indeed, effective $\bar{I}$-induced vertex (5.28)
is Q-invariant, so the BRST-symmetry is broken by the choice of the
vacuum state at finite $\mid v\mid $. However, the appearance of UV
cut-off parameter in our results for correlation function (\ref{anw})
and vacuum energy (6.2) signals that the non-perturbative conformal
invariance breaking occurs (in addition to the perturbative one,
associated with the non-zero $\beta $-function).
The latter is perhaps related to some new anomaly associated with the
non-compactness of the target space. The situation could be similar
to that with chiral anomaly in 4D Gauge theories. In the latter case
instantons do not produce the chiral symmetry breaking by themselves.
They just saturate the chiral anomaly relation.

Now let us discuss the question of the lifetime of states with finite
 $\mid v \mid $.
 We give a qualitative arguments
below that it is infinite. The decay rate of an unstable vacuum state is
proportional to \cite{Col}
\beq
\Gamma\sim e^{-S_{b}},
\eeq
where $S_{b}$ is the action on the bubble of the true vacuum inside the
false one. In order to estimate $\Gamma $, one can try to pick up some bubble
configuration $w_{b} $ which goes to infinity at some point $x_{0}$,
while at large $\mid x-x_{0} \mid $ approaches finite VEV v.

Observe now, that instanton (2.19) by itself plays a role of such a bubble.
However, it is a stable solution of equations of motion and has no negative
modes. Therefore it can not give a contribution to the decay rate
of the false vacuum. Furthermore, consider the $I\bar{I}$  pair.
Sometimes in nontopological theories $I\bar{I} $-induced vacuum energy could
have an imaginary part which reflects the non-borel summable behavior of
the perturbation theory \cite{Zinn}. However, we got no imaginary part in
(6.2) (at least to the leading order in $log L/a$). That means that $\Gamma $
in
(7.1) is suppressed at least as $1/log L/a$. Thus we can live forever in a
broken phase of TFT in the toy model we consider in this paper and never
realize that it is unstable.

Let us note, however that our calculation of the $I\bar{I}$ vacuum energy in
Sect.6 could be viewed only as a preliminary estimate. The point is that the
answer in (6.2) contains the IR logarithmic divergence. It actually arises
from the integral over the size of the instanton (the size of $\bar{I}$
is small $\mid \r _{a} \mid \sim a $ ). Similar infrared divergences
(related to the integration of the $I$ size ) is known to occur in
$O(3) $ $\sigma $-model \cite{FFS}. In $O(3)$ $\sigma $-model each $I $
can be represented as a dipole of some ``charge'' and ``anticharge''
\cite{FFS}. The infinitely growing size of $I $ then means the phase
transition (of the Kosterlitz-Thouless type ) from the dipole phase of
these ``charges'' to the plasma one. Something similar could happen
in the model at hand. The issue needs a future study. If the above mentioned
phenomenon happened in the black hole topological $\sigma $-model,
then the scalar potential and the vacuum energy could be different from
the one in (6.2) (still nonzero, of course).

Another question which remains unsolved in this paper is: what kind of
physics emerge in the broken phase of the topological
$\sigma $ model at hand. The result in (4.4) shows that some physical degrees
 of freedom are ``liberated''. However, it is unclear if (4.4) could be
interpreted as an exchange of one or two boson particles or not.
To clarify this question one has to study multipoint correlation functions
(2.14) which receive contributions of instantons with $k>1$. It is also
clear, that the $\bar{I} $ vertex in (5.28) gives rise to a nontrivial
interactions, which would be interesting to study in more details.

Another intriguing direction of thinking is the possible string theory
applications of the breakdown of the BRST-symmetry we find in the present
paper. In particularly, in ref.\cite{MV} the $c=1 $ string is shown to be
equivalent to topological Kazama-Suzuki $SL(2,R)/U(1) $ coset. On the top
of this it is interesting to consider the possible instanton solutions in
$SL(2,R)/U(1) $ gauged $ WZW $ theory. It differs from the model we study
in this paper by the presence of the dilaton term. As dilaton term is a
quantum correction one could expect that its presence cannot change the
fact of the existence of instantons in the model. Then the emergence of
instantons could produce dramatic consequences for the string theory.
Of course, if the conformal invariance of 2D theory is broken it
cannot serve as a string vacuum state any longer. However, if we think of
quantum string theory, we might have to consider these states as well.
This point of view has been recently taken up in refs.\cite{JE}.
In latter papers the renormalization group flow (which could occur in
certain black hole $ \sigma$-models if instantons are taken into account )
is interpreted as a decay of the false string vacuum and related to the
black hole information loss paradox \cite{Haw}. If the phenomenon we observe
in this paper is indeed related to some kind of anomaly ( as we conjecture
above ) than it could appear to be an unavoidable feature of the string in the
 black hole background.

The author is grateful to D. Amati, P. Damgaard, A. Johansen, A. Losev,
 N. E. Mavromatos, A. Schwimmer
for stimulating discussions and to the Particle Physics Group at the
University of Wales, Swansea where  part of this work was done for the
hospitality. This work was supported in part by the Royal Society
Kapitza fellowship and Higher Education Funding Council for Wales.

\end{document}